%% file: main.tex
\documentclass[journal]{IEEEtran}

\usepackage[pdftex]{graphicx}
\usepackage{amsmath}
\usepackage{array}
\usepackage{bm}
\usepackage{amsthm}
\usepackage{amsfonts}
\usepackage{amssymb}
\usepackage{mathrsfs}
\usepackage{cite}
\usepackage{wrapfig} 
\usepackage{indentfirst}
\usepackage{multirow}
\usepackage{color}
\usepackage{graphicx}
\usepackage{mathtools}
\usepackage[ruled]{algorithm2e}
\usepackage[export]{adjustbox}
\usepackage[flushleft]{threeparttable}

\DeclareGraphicsExtensions{.pdf,.jpeg,.png}
\begin{document}

\title{Penalized PET/CT Reconstruction Algorithms with Automatic Realignment for Anatomical Priors}

\author{Yu-Jung~Tsai, Alexandre Bousse, Simon Arridge, Charles W. Stearns,~\IEEEmembership{Fellow,~IEEE}, Brian F. Hutton,~\IEEEmembership{Senior Member,~IEEE} and Kris~Thielemans,~\IEEEmembership{Senior Member,~IEEE}
\thanks{This work was supported in part by GE Healthcare and in part by the National Institute for Health Research, University College London Hospitals Biomedical Research Centre. \emph{Asterisk indicates corresponding author.}}

\thanks{Y.-J. Tsai$^\ast$ was with the Institute of Nuclear Medicine, University College London, London NW1~2BU, UK and is now with the Yale PET Center, Yale University, New Haven, CT 06520, USA. (e-mail: yu-jung.tsai@yale.edu).}
\thanks{A. Bousse was with the Institute of Nuclear Medicine, University College London, London, UK and is now with the LaTIM, INSERM, UMR 1101, Universit\'e de Bretagne Occidentale, Brest, France.}
\thanks{S. Arridge is with the Department of Computer Science, University College London, London WC1E~6BT, UK.}
\thanks{C. W. Stearns is with MICT Engineering, GE Healthcare, Waukesha, WI~53188 USA.}
\thanks{B. F. Hutton is with the Institute of Nuclear Medicine, University College London, London, UK.}
\thanks{K. Thielemans is with the Institute of Nuclear Medicine, University College London, London NW1~2BU, UK.}}

\maketitle

\begin{abstract}
Two algorithms for solving misalignment issues in penalized PET/CT reconstruction using anatomical priors are proposed. Both approaches are based on a recently published joint motion estimation and image reconstruction method. The first approach deforms the anatomical image to align it with the functional one while the second approach deforms both images to align them with the measured data. Our current implementation alternates between alignment estimation and image reconstruction. We have chosen Parallel Level Sets (PLS) as a representative anatomical penalty, incorporating a spatially-variant penalty strength. The performance was evaluated using simulated non-TOF data generated with an XCAT phantom in the thorax region. We used the attenuation map in the anatomical prior. The results demonstrated that both methods can estimate the misalignment and deform the anatomical image accordingly. However, the performance of the first approach depends highly on the workflow of the alternating process. The second approach shows a faster convergence rate to the correct alignment and is less sensitive to the workflow. The presence of anatomical information improves the convergence rate of misalignment estimation for the second approach but slow it down for the first approach. Both approaches show improved performance in misalignment estimation as the data noise level decreases.
\end{abstract}

\input{intro}

\input{method}

\input{evaluation}

\input{results}
\input{discussion}
\input{conclusion}
\bibliographystyle{IEEEtran}
\bibliography{ref}

\end{document}

%% file: intro.tex
\section{Introduction}
Penalized maximum-likelihood (PML) image reconstruction using penalties derived from anatomical images, such as computed tomography (CT) or magnetic resonance (MR) images, has been shown to be effective in improving object delineation and reducing quantitative error in many studies \cite{Vunckx2012,Tsai2015,Sakaguchi2008,Gindi1993,Lipinski1997,Comtat2002,Mameuda2007,Nuyts2005,Bruyant2004,Browne1996}. However, in order to utilize the structural information without incurring artifacts, a good alignment between the anatomical and functional images is essential \cite{Comtat2002,Bruyant2004,Kulkarni2007,Ehrhardt2014,Deidda2019}. This is challenging in practice because these images are most likely obtained separately or sequentially. Even with a multi-modality scanner that performs simultaneous acquisition (e.g., Siemens mMR system and GE Signa PET/MR), good alignment between the acquisitions is still difficult to achieve due to the different time scales of the scans. For the misalignment involving non-rigid deformation, for instance due to respiration, the assumption is even harder to satisfy and accurate image segmentation and co-registration (which are difficult and time consuming) are often required \cite{Liao2007,Camara2007}. 

In many cases, anatomical information can be derived from the attenuation image. Since the literature on solving the misalignment issue between a priori anatomical information and the functional information is quite limited, we instead seek ideas from a similar but previously studied problem in thoracic PET imaging, in which a potentially misaligned CT or MR-derived attenuation map is used for the attenuation correction. As for PML image reconstruction using anatomical priors, the misalignment induced by patient respiration degrades resolution of the reconstructed image and introduces artifacts where large movement or deformation of organs is observed \cite{Xu2011,Nyflot2015}. Although these methods are intended to be used for having a better attenuation corrected image from emission data, they offer insights into resolving the misalignment between the anatomical attenuation map and the functional emission image. 

This study will concentrate on imaging of the thorax, where respiratory motion is a known problem \cite{pepin2014}. One strategy to tackle the respiratory motion is to sort the acquired data from both modalities into several gates where no motion is assumed in each of them. The gated data are then paired up according to their breathing phases estimated from the data themselves or an external tracking system \cite{Nehmeh2008}. In addition to reconstructing these gated data pairs individually and then registering them to a reference respiratory phase \cite{Nehmeh2004}, one can also incorporate the corresponding attenuation information into 4-dimensional (4-D) reconstruction algorithms \cite{Manjeshwar2006,Qiao2006}. However, these methods rely on a relatively consistent breathing pattern during both scans \cite{Teo2012}. Moreover, they imply the need of special scans to obtain the gated anatomical information. This can increase patient dose or prolong the overall study time, depending on the applied anatomical imaging modality. To adapt to irregular breathing patterns, another strategy applies individual motion model to deform the input attenuation map \cite{McClelland2013}. The model can be derived from other imaging data, such as dynamic CT and MR \cite{Manber2016}, as well as the non-attenuation corrected PET \cite{Kalantari2017}. However, the former approach has the potential problem of propagating the error in the model estimation to the final reconstructed image, while the latter method's performance depends on the tracer distribution and data statistics. Another way around these issues is to use population-based deformation models \cite{McQuaid2011,Fayad2009}. However, these have not been convincingly shown to work in practice \cite{McClelland2013}. The application of applying an individual model to PML image reconstruction using anatomical priors therefore faces similar challenges.
Finally, none of the above methods is able to cope with the residual misalignment caused by other general motion of the patient.

Algorithms that allow simultaneous estimation of the activity distribution and the corresponding attenuation map from the respiratory gated PET data have been proposed \cite{Nuyts1999,Rezaei2012} in recent years. These methods do not rely on assumptions about the breathing pattern or a pre-estimated motion model. Therefore, they have the potential to be applied to different misalignment problems without suffering from the error propagation issue. However, since the problem is very ill-conditioned, some \textit{a priori} knowledge about the intensity distribution of the attenuation map is required. This can compromise the benefit of using anatomical information during the image reconstruction as the intensity is restricted to several values and most of the anatomical details are lost. Besides, significant cross-talk between the estimated activity and attenuation map is observed in non-time-of-flight (non-TOF) PET. Although the artifacts can be largely eliminated when TOF data are available, the \textit{a priori} knowledge about the intensity distribution is still necessary \cite{Rezaei2012}.

In contrast to seeking to align the attenuation map with the emission image, a different joint estimation approach incorporates a warp matrix that deforms both the activity distribution and the attenuation map within the objective function \cite{Bousse2016,Rezaei2016,Bousse2017}. By optimizing the objective function using an alternating process between motion estimation and image reconstruction, the motion compensation and attenuation correction are achieved simultaneously. The optimization can be applied to both non-TOF and TOF data albeit with a significantly improved convergence rate when TOF data are available\cite{Bousse2016b}. The \textit{a priori} knowledge of the attenuation distribution is not necessary anymore. This study motivated us to investigate the idea of applying a warp matrix to an anatomical prior. Extending on the existing method, two approaches that account for the misalignment between the functional and anatomical images by incorporating the warp matrix into the penalized objective function are proposed. The proposed methods do not require any image preprocessing, such as segmentation and co-registration. They are also applicable to both non-TOF and TOF datasets and no \textit{a priori} information regarding the misalignment or intensity distribution of both images are needed. To the best of our knowledge, there are no other algorithms that show similar capacities in literature. As a special case of the application, we will only investigate the alignment of one PET position with a single CT derived attenuation map, which is also used to provide anatomical information. This paper is an extension of initial results presented in \cite{Tsai2018}.

%% file: method.tex
\section{method}
\subsection{Objective function without considering the potential misalignment}
In this section, we define the PML objective function with an anatomical prior without misalignment considerations between the activity and anatomical images. Given the emission image $\bm{f}\in\mathbb{R}^J$, the anatomical image $\bm{z}\in\mathbb{R}^J$, the attenuation map $\bm{\mu}\in\mathbb{R}^J$ and the measured data $\bm{g}\in\mathbb{R}^I$, the objective function can be written as: 
\begin{equation}\label{PML}
	\Phi(\bm{f}) = -L(\bm{f},\bm{g},\bm{\mu}) + \beta R(\bm{f}|\bm{z})\,,
\end{equation}
where $L$ is the log-likelihood and $R$ is the penalty function with a parameter $\beta$ controlling its strength. 

As the statistical nature of the measured data $\bm{g}$ can be described using the Poisson distribution in emission tomography, the log-likelihood function $L$, omitting terms independent of $\bm{f}$, can be expressed as:
\begin{align}\label{likelihood}
L(\bm{f}, \bm{g}, \bm{\mu}) &= \sum_i g_i \log \bar{g}_i(\bm{f}, \bm{\mu}) - \bar{g}_i(\bm{f}, \bm{\mu})\,, \nonumber \\
\bar{\bm{g}}(\bm{f}, \bm{\mu}) &\triangleq \mathrm{diag}\left\{e^{-\bm{A}\bm{\mu}}\right\} \bm{A}\bm{f} + \bm{n}
\end{align}  
where $\bar{g}_i$ is the mean measurement in bin $i$, $\bm{A}\in\mathbb{R}^{I\times J}$ is the system matrix which characterizes the physical system properties, such as resolution and detector sensitivity, in terms of detection probability and $\bm{n}\in\mathbb{R}^I$ is the expected background events vector. The attenuation effect is modeled explicitly by the matrix $\mathrm{diag}\left\{e^{-\bm{A}\mu}\right\}$, where $\mathrm{diag}\left\{\cdot\right\}$ is an operator that constructs a diagonal matrix from a vector.

Since it has shown promising results in the literature \cite{Ehrhardt2014,Ehrhardt2015,Ehrhardt2016}, Parallel Level Sets (PLS) is chosen as a representative anatomical penalty in this study:
\begin{align} 
R(\bm{f}|\bm{z}) &= \sum_j \sqrt{\epsilon^2 + \|\left[\nabla\bm{f}\right]_j\|_2^2 - {\langle \left[\nabla\bm{f}\right]_j, \left[\bm{\xi}\right]_j\rangle}^2}, \nonumber \\
\left[\bm{\xi}\right]_j &\coloneqq \frac{\left[\nabla\bm{z}\right]_j}{\sqrt{\|\left[\nabla\bm{z}\right]_j\|_2^2 + \eta^2}},~\epsilon~\mathrm{and}~\eta > 0 
\end{align}
where $\nabla$ is the gradient operator, $\langle\cdot,\cdot\rangle$ is the Euclidean scalar product and $\|\cdot\|_2$ denotes the $\ell^2$-norm. The edge preserving property of PLS is modulated by the pair of parameters $(\epsilon,\eta)$. 

\subsection{Objective function considering the potential misalignment}\label{JRM}
Two approaches that account for the misalignment between the functional and anatomical images in penalized image reconstruction using anatomical priors are proposed in this section. Both approaches are based on a joint motion estimation and image reconstruction method proposed recently for dealing with the mismatch between the attenuation map and the PET image in respiratory gated PET/CT \cite{Bousse2016}. Instead of applying a quadratic penalty function to enforce smoothness on the reconstructed activity images as in \cite{Bousse2016}, an anatomical penalty calculated with the attenuation map is employed to improve the image quality. Therefore, artifacts induced by the misalignment between the activity image and the attenuation map will be introduced through both the attenuation correction and the incorporated penalty function. The main difference between these two approaches is that the first approach $\Phi_1$ (Approach I) deforms the anatomical image (\textit{i.e.}, the attenuation map) to align it with the functional image, while the second approach $\Phi_2$ (Approach II) deforms both images to align them with the measured data. Although the deformation of the attenuation map implies the change of the scatter distribution, the estimated background events are fixed during the optimization process for simplicity. As in \cite{Bousse2016}, we use uniform cubic B-splines for image interpolation and deformation. Despite the use of an anatomical prior, Approach II is quite similar to the method introduced in \cite{Bousse2016} and Approach I is essentially a simplified algorithm to Approach II. We also modified Approach II to impose positivity on image values as opposed to B-spline coefficients. 

Assume a continuous volumetric image $s:\mathbb{R}^3\rightarrow\mathbb{R}$ can be represented as a linear combination of basis functions centered on a voxel grid $\mathsf{C} = \{\bm{r}_k,~k = 1, \dots,~N\}\subset\mathbb{R}^3$ that coincides with the voxel centers:
\begin{equation}
s(\bm{r}) = \sum_{k=1}^N s_k' B\left(\frac{\bm{r}-\bm{r}_k}{\sigma_1}\right)\,,
\end{equation}
where $s_k'$ is the B-spline coefficient of the basis function centered on voxel $k$, $\bm{r} = (x, y, z)$ is the index vector in the 3-D Cartesian coordinate system, $B(\bm{r}) = b(x)b(y)b(z)$ is an interpolating function based on the cubic B-splines $b$ and $\sigma_1$ represents the voxel-spacing. The discretized image can therefore be represented as a collection of the B-spline coefficients $\bm{s}' = (s_j')_{j=1}^{N}$. From this section, the prime notation is used to distinguish the B-spline coefficients from the corresponding voxel values for images. Note that the coefficients of the cubic B-splines are not identical to the image values at the grid nodes. Particularly, the B-spline coefficients can be negative. The deformation of the image represented by the coefficients $\bm{s}'$ is achieved by deforming the continuous image function $s$ followed by a re-sampling on $\mathsf{C}$ for every voxel $j$:
\begin{equation}\label{image_warp}
[\bm{W}\bm{s}']_j = \sum_{k=1}^N s_k' B\left(\frac{\nu(\bm{r}_j)-\bm{r}_k}{\sigma_1}\right)\,,
\end{equation}
where $\bm{W}$ is a square matrix with each element $[\bm{W}]_{j, k} \triangleq B(\frac{\nu(\bm{r}_j)-\bm{r}_k}{\sigma_1})$ and $\nu$ is the warping function. Given $\tilde{\mathsf{C}} = \{\tilde{\bm{r}}_l,~l = 1, \dots,~\tilde{N}\}$ a uniform sub-grid of $\mathsf{C}$ with $\tilde{N}$ grid nodes, the function $\nu_\theta$ can be parametrized by a collection of the deformation coefficients $\bm{\theta} = (\bm{\theta}^x,~\bm{\theta}^y,~\bm{\theta}^z)$:
\begin{equation}
\nu_\theta(\bm{r}) \triangleq \bm{r} + \left[
\begin{aligned}
\sum_{l=1}^{\tilde{N}} \theta_l^x B(\frac{\bm{r}-\tilde{\bm{r}}_l}{\sigma_2}) \\
\sum_{l=1}^{\tilde{N}} \theta_l^y B(\frac{\bm{r}-\tilde{\bm{r}}_l}{\sigma_2}) \\
\sum_{l=1}^{\tilde{N}} \theta_l^z B(\frac{\bm{r}-\tilde{\bm{r}}_l}{\sigma_2})
\end{aligned}
\right] = \left[
\begin{aligned}
\nu^x(\bm{r}) \\
\nu^y(\bm{r}) \\
\nu^z(\bm{r})
\end{aligned}
\right]
\end{equation}
where $\sigma_2$ is the distance between two grid nodes.

\subsubsection{Approach I}
The first approach optimizes an objective function $\Phi_1$ that considers the deformed anatomical image. Assume the attenuation map $\bm{\mu}$, represented as a collection of the B-spline coefficients $\bm{\mu}'=(\mu_j')_{j=1}^N$, is used to provide anatomical information as well, $\Phi_1$ is given by:
\begin{align}\label{approach1}
	\Phi_1(\bm{f}, {\bm{\mu}'},\bm{\theta}) &= -L(\bm{f}, \bm{g}, \bm{W}{\bm{\mu}'}) + \beta R(\bm{f}|\bm{W}{\bm{\mu}'}) \\ \nonumber
	&+ \gamma U(\bm{\theta})\,, 
\end{align}
where $U(\bm{\theta})$ is a quadratic penalty on the difference between neighboring nodes of the motion grid for reducing the influence of noise and $\gamma$ is a constant that controls its strength. Note that the misalignment between $\bm{\mu}$ and $\bm{f}$ affects the optimization through the attenuation correction in the log-likelihood $L$ and the incorporated anatomical penalty $R$ as both functions use the warped attenuation map $\bm{W}{\bm{\mu}'}$ as inputs. Since this approach does not require deforming the activity image, $\bm{f}$ in \eqref{approach1} represents a vector of image values. The positivity constraint on $\bm{f}$ can therefore be achieved by performing constrained image reconstruction. In contrast, the attenuation map is represented as a collection of coefficients ${\bm{\mu}'}$ for image warping using B-splines. This could lead to negative values in $\bm{W}{\bm{\mu}'}$ as we are optimizing the B-spline coefficients. However, since small negative values in $\bm{W}{\bm{\mu}'}$ would become attenuation factors very close to one, they have been left unchanged in this study. When the objective function in \eqref{approach1} is optimized, the warped attenuation map should be in the same space as the activity image and $e^{-\bm{AW\mu}}$ is a coefficient vector that accurately corrects $\bm{f}$ for attenuation in the projection domain. In other words, $\bm{W}\bm{\mu}'$ and $\bm{f}$ are aligned and adequately projected to fit the data $\bm{g}$.

\subsubsection{Approach II}
Instead of seeking to align the warped attenuation map with the reconstructed activity image, the second approach deforms both the anatomical and functional images in order to obtain an estimate that optimizes the objective function $\Phi_2$:
\begin{align} \label{approach2}
	\Phi_2({\bm{f}'}, {\bm{\mu}'},\bm{\theta}) &= -L(\bm{W}{\bm{f}'}, \bm{g},\bm{W}{\bm{\mu}'}) \!+\! \beta R(\bm{f}|\bm{\mu}) \\ \nonumber
	&+ \gamma U(\bm{\theta}) \!+\! \delta E({\bm{f}'})\,,
\end{align}
where
\begin{equation}
E(\bm{s}) = \sum_{n=1}^{\hat{N}}\mathrm{min}(0,~s(\nu_\theta(\hat{\bm{r}}_n)))^2
\end{equation}
is a barrier function that penalizes negative values in $\bm{W}{\bm{f}'}$ \cite{Nichols2002}. Given $\hat{\mathsf{C}} = \{\hat{\bm{r}}_n,~n=1,\dots,~\hat{N}\}$ a finer grid that contains a finite number of uniformly spaced locations in each interval of two grid nodes in the voxel grid $\mathsf{C}$, the function computes the spline values (\textit{i.e.}, the image values) centered on the finer grid and penalizes the square of any negatives. The strength of $E$ is determined by the parameter $\delta$. In this study, we defined the distance between two adjacent locations in $\hat{\mathsf{C}}$ equal to one quarter of the grid spacing used for the image. \\
Although the log-likelihood requires the warped emission and anatomical images as inputs, the anatomical penalty function is calculated with the image value of the non-warped ones. Since the motivation of using an anatomical prior is to encourage edges in the emission image corresponding to those in the anatomical one, finding common edges in the warped or non-warped images is essentially a similar optimization problem. In other words, calculating the anatomical penalty function with $\bm{W}{\bm{f}'}$ and $\bm{W}{\bm{\mu}'}$ should lead to (nearly) the same solution. \\
In summary, when the objective function in \eqref{approach2} is optimized, the emission and anatomical images are aligned such that the attenuation corrected projections calculated with $\bm{W}\bm{f}'$ and $\bm{W}\bm{\mu}'$ fit the measured data $\bm{g}$.  

\subsection{Algorithm implementation}\label{AlgoImpt5}
In both approaches the optimization was implemented as an alternating process that includes a misalignment estimation subroutine and a penalized image reconstruction subroutine. Pseudo-code that summarizes the implementation can be found in Algorithm~\ref{algo3}. The workflow was defined by the number of inner iterations (InnerIter1 and InnerIter2 in Algorithm~\ref{algo3}) for these two subroutines and the number of outer iterations (OuterIter in Algorithm~\ref{algo3}) that controls the repetition of the alternating process. Since limited-memory Broyden-Fletcher-Goldfarb-Shanno (L-BFGS) based algorithms show the potential to work with a relatively wide range of penalty functions \cite{Tsai2019}, we applied L-BFGS for unconstrained optimization (misalignment estimation in both approaches and image reconstruction in the second approach) and L-BFGS-B \cite{Byrd1995} for the positivity constrained image reconstruction in Approach I. The statement $x \leftarrow$ L-BFGS$(G, x_0, \mathrm{Niter})$ in Algorithm~\ref{algo3} means optimizing function $G$ using algorithm L-BFGS with initial point $x_0$ and $\mathrm{Niter}$ iterations, where $x$ can be either activity image or deformation coefficients depending on the context. Ordered subsets expectation maximization (OS-EM) was used to reconstruct the initial activity image $\bm{f}_0$ of the whole process. The misalignment estimation was then initialized by $\bm{f}_0$ and the attenuation map $\bm{\mu}$. The implementation of the misalignment estimation employed in this study was originally proposed in \cite{Bousse2016}. Every time the misalignment estimation was done, a new initial image $\bm{f}_\mathrm{InitInner}$ for the penalized image reconstruction was recomputed using OS-EM, taking into account the current estimated misalignment. To improve the convergence rate of the penalized image reconstruction, a preconditioner proposed in \cite{Tsai2018b} was incorporated into both approaches. A spatially-variant penalization scheme was also applied to the anatomical penalty function to further achieve uniform local contrast \cite{Tsai2019}. Both the preconditioner and the spatially-variant penalty strength were calculated with the initial image from OS-EM at every outer iteration as well.
\begin{algorithm}
\small
\SetAlgoLined
\KwIn{Data $\bm{g}$, attenuation map $\bm{\mu}$ and strength of each penalty function (\textit{i.e.}, the set of parameters ($\beta, \gamma$) for Approach I and ($\beta, \gamma, \delta$) for Approach II}
\KwOut{Estimated tracer distribution $\bm{f}$ and B-spline deformation coefficient $\bm{\theta}$}
$\bm{\theta}_0 \leftarrow \bm{0}$ \;
$\bm{f}_0 \leftarrow$ OS-EM$(\bm{g},\bm{\mu},\bm{\theta}_0)$ \;
\For{$t = 0, \ldots, \mathrm{OuterIter}-1$}{
\textit{misalignment estimation}\;
\uIf{$\mathrm{Approach~I}$}{Define $F \colon \bm{\theta} \mapsto \Phi_1(\bm{f}_t, \bm{\mu}, \bm{\theta})$ \;}
\uElseIf{$\mathrm{Approach~II}$}{Define $F \colon \bm{\theta} \mapsto \Phi_2(\bm{f}_t, \bm{\mu}, \bm{\theta})$\;}
{\textbf{end}\\}
$\bm{\theta}_{t+1} \leftarrow$ L-BFGS $(F, \bm{\theta}_t$, InnerIter1)\;
\textit{image reconstruction}\;
$\bm{f}_\mathrm{InitInner} \leftarrow$ OS-EM$(\bm{g},\bm{\mu},\bm{\theta}_{t+1})$ \;
\uIf{$\mathrm{Approach~I}$}{Define $H \colon \bm{f} \mapsto \Phi_1(\bm{f}, \bm{\mu}, \bm{\theta}_{t+1})$ \;
$\bm{f}_{\mathrm{InnerIter2}} \leftarrow$ L-BFGS-B$(H, \bm{f}_\mathrm{InitInner}$, InnerIter2) \;}
\uElseIf{$\mathrm{Approach~II}$}{Define $H \colon \bm{f} \mapsto \Phi_2(\bm{f}, \bm{\mu}, \bm{\theta}_{t+1})$ \;
$\bm{f}_{\mathrm{InnerIter2}} \leftarrow$ L-BFGS$(H, \bm{f}_\mathrm{InitInner}$, InnerIter2) \;}
{\textbf{end}\\}
$\bm{f}_{t+1} \leftarrow \bm{f}_{\mathrm{InnerIter2}}$ \;
}
\caption{Pseudo-code for the alternating process}\label{algo3}
\end{algorithm}

%% file: evaluation.tex
\section{Evaluation}
Since both approaches should be able to find a warped attenuation map that helps minimize the objective function, the evaluation is focused on the performance of the misalignment subroutine of each approach.

\subsection{Data}
\begin{figure}
	\centering
	\includegraphics[width=0.49\linewidth]{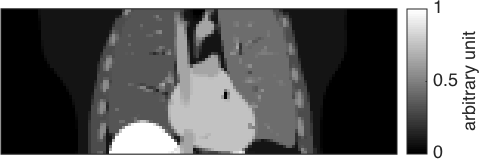}
    \includegraphics[width=0.49\linewidth]{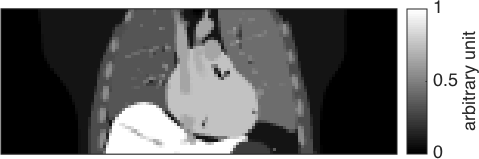}
    \includegraphics[width=0.49\linewidth]{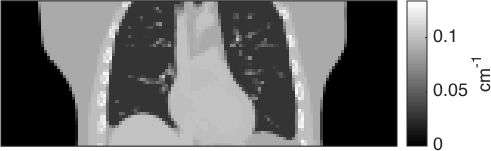}
    \includegraphics[width=0.49\linewidth]{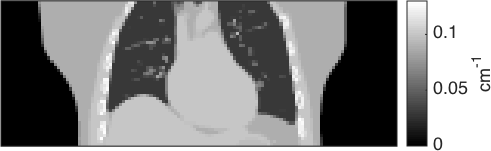}
    \caption{The central coronal view of the XCAT phantoms representing end inspiration (top left) and expiration (top right). The corresponding $\bm{\mu}$ maps are also provided (bottom).}
    \label{fig:fig1}
\end{figure}

Two XCAT phantoms \cite{Segars2010} representing different respiratory phases and the corresponding $\bm{\mu}$ maps (Figure~\ref{fig:fig1}) were produced to simulate different PET positions. Both phantoms were a $128 \times 128 \times 47$ matrix with voxel size of $3.906$ mm in all directions. The set of images at end inspiration was forward projected in parallel acquisition geometry to generate data with size of $128$ bins $\times~140$ angles $\times~47$ slices. A uniform projection with a constant intensity was added to the generated data to simulate the background events equivalent to $78.66\%$ of the total prompts. Both approaches were initially evaluated with a noiseless dataset. To further assess performance in the presence of noise, three datasets with total counts of $16$~M, $54$~M and $161$~M were also simulated using a built-in Poisson noise model in MATLAB. The same as for the noiseless dataset, uniform background events were simulated and added to the generated sinograms. The amount of background events was equivalent to $78.66\%$ of the total prompts for each noisy dataset. Note that all simulations took into account the attenuation effect and system blurring using FWHM = $5$ mm in all directions. To simulate the misalignment between functional and anatomical images, the attenuation map at end expiration was used as the initial input for both attenuation correction and misalignment estimation.

\subsection{Reconstruction}
The selection of the strength of each penalty  was based on an initial investigation, where the difference between the warped and target $\bm{\mu}$ maps at OuterIter = $30$ was visually compared with respect to a given set of candidate strengths (results not shown). A simple grid search using relatively large magnitude differences between consecutive candidates ($10$ fold for $\beta$ and $\delta$ (when applicable), and $100$ fold for $\gamma$) was performed on the noiseless dataset. The combination of penalty parameters that resulted in the best visual alignment in the diaphragm and heart regions was recorded. As a result, for the noiseless dataset, the set of parameters that determines the strength of each penalty function was $(\beta, \gamma) = (7\times 10^{-3}, 10^{-4})$ for $\Phi_1$ and $(\beta, \gamma, \delta) = (7\times 10^{-3}, 10^{-4}, 10^{-1})$ for $\Phi_2$. A stronger $\beta = 2\times10^{-1}$ was used when data with noise were considered for either approach. The strength of other penalty function(s) remained the same as for the noiseless dataset. The parameter $\epsilon$ in PLS was fixed at $10^{-1}$ for the noiseless data. To have similar local effect of PLS for data with different noise level, $\epsilon$ was chosen according to the scale of the statistics to reconstruct the activity images. Therefore, it was $10^{-2}$, $3\times10^{-2}$ and $10^{-1}$ for data with total counts of $16$~M, $54$~M and $161$~M, respectively. As the scale of the anatomical image, \textit{i.e.}, the attenuation map $\bm{\mu}$, did not vary with data noise level, $\eta = 10^{-2}$ was used for all datasets. The distance between two grid nodes for the deformation model was $6$ voxels. The alternating process as well as the image reconstruction subroutine at every outer iteration were initialized by one full iteration of OS-EM with $14$ subsets. This initial image was also used for calculating the preconditioner and the spatially-variant penalty strength for the anatomical prior. Up to $100$ outer iterations were used for both approaches. Each reconstructed activity image and the warped attenuation map had the same matrix dimension and voxel size as that of the phantoms. 

\subsection{Analysis}
The influence of the workflow configuration on the performance of misalignment estimation of each approach was studied with both noiseless and noisy data. For the noiseless data, a two-part study was performed. In the first part of the study, we used $1$ inner iteration for the misalignment estimation and explored the minimum iterations required for the image reconstruction subroutine to obtain satisfactory results. The studied inner iteration numbers for the image reconstruction subroutine (InnerIter2) were $1$, $5$, $10$ and $20$. We then fixed the iteration number for the image reconstruction to the optimal value found in the first part and increased the number of iterations for the misalignment estimation (InnerIter1) from $1$ to $5$, $15$ or $30$ to assess if the performance of the misalignment estimation can be improved by using a higher inner iteration number. Up to $100$ outer iterations (controlling the repetition of the alternating process) were performed. A workflow that provided satisfactory visual alignment in the diaphragm and heart regions for both approaches was also noted such that their performance at the same computational cost could be compared. This common workflow was then also used to assess if the use of an anatomical prior is beneficial for either approach. Reconstructions without considering any anatomical information were achieved by substituting the anatomical image (\textit{i.e.}, $\bm{\mu}$ map at end expiration) with a uniform image when calculating PLS, therefore it is equivalent to using a (smooth) total variation (TV) prior.

The results for the noiseless data formed the basis for the investigation on noisy datasets. We started with seeking for a common workflow for both approaches using the dataset with total counts of $161$~M. Recall that a stronger $\beta$ was used for the anatomical penalty function in the presence of noise. As the convergence rate of the preconditioned L-BFGS-B (L-BFGS-B-PC) varies with the strength of the penalty function and data noise level \cite{Tsai2018b}, the alternating process was also performed with a higher InnerIter2 = $20$ or $40$ for both approaches. The number of outer iterations and the number of inner iterations for the misalignment subroutine were kept the same as for the noiseless dataset. We evaluated the value of InnerIter2 that was able to provide satisfactory results at $100$ outer iterations for both approaches and used it to determine the common workflow. This workflow was then applied to all noisy datasets. We did not seek for the optimal common workflow for each noise level; this ensured that the sensitivity of the approaches to noise was explored without interferences from changing of workflow.

The difference image between the warped and target $\bm{\mu}$ maps at a given outer iteration was used to evaluate the performance of misalignment estimation. As the lungs are the main target of the respiratory motion alignment, a mask slightly larger than the lungs was applied to the difference images and the root-mean-square errors (RMSE) was computed in the mask to reflect the misalignment estimation of borders and small structures of the lungs. The yellow contour in the top-left image of Fig.~\ref{fig:fig2} illustrates the edges of the mask in the central coronal view. The RMSE in the mask was then plotted against the outer iteration numbers to assist evaluations of the performance change of each approach in misalignment estimation in response to the change of workflow configuration, availability of anatomical information and data noise level.

%% file: results.tex
\section{Results}
\subsection{Influence of workflow configuration (noiseless data)}\label{best_work_flow}
\begin{figure}
	\centering
	\includegraphics[width=0.49\linewidth]{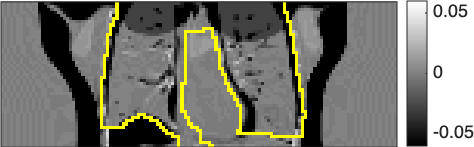}
	\includegraphics[width=0.49\linewidth]{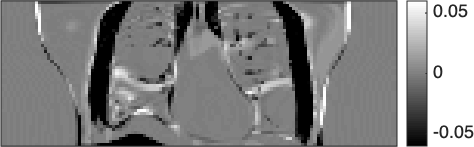}
	\includegraphics[width=0.49\linewidth]{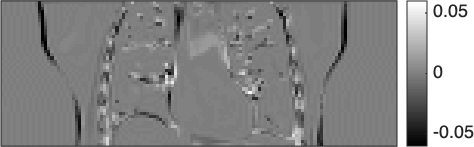}
	\includegraphics[width=0.49\linewidth]{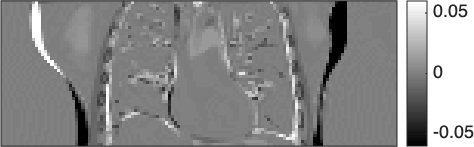}
	\includegraphics[width=0.49\linewidth]{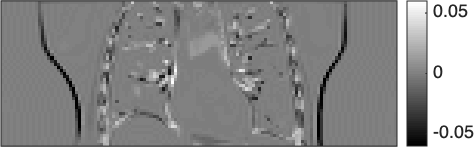}
	\includegraphics[width=0.49\linewidth]{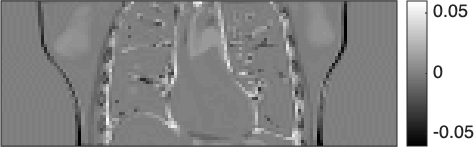}
	\includegraphics[width=0.49\linewidth]{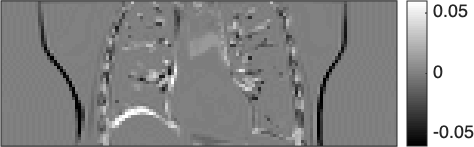}
	\includegraphics[width=0.49\linewidth]{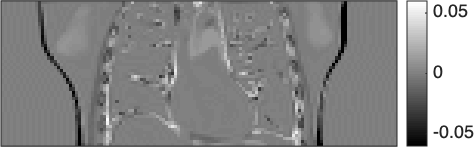}
	\caption{The central coronal view of the difference images between the target and $\bm{W}{\bm{\mu}'}$ maps for Approach I (left column) and Approach II (right column) at $100$ outer iterations. The applied workflows were $1$ inner iteration for the misalignment estimation and $1$ (top row), $5$ (second row), $10$ (third row) and $20$ (bottom row) inner iterations for the image reconstruction. The noiseless dataset was used. The yellow contour in the top-left image illustrates the edges of the mask in the central coronal view.}
	\label{fig:fig2}
\end{figure}

For the sub-study in which the inner iteration number for the misalignment estimation was fixed at $1$, the central coronal view of the difference images between the warped and target $\bm{\mu}$ maps at OuterIter = $100$ are shown in Fig.~\ref{fig:fig2}. For both approaches, using InnerIter2 = $1$ was problematic, resulting in severe distortion of structures in the warped $\bm{\mu}$ map (Fig.~\ref{fig:fig2}, top row). Satisfactory results were obtained with Approach I when InnerIter2 = $5$ or $10$ was chosen (Fig.~\ref{fig:fig2}, left column, second and third rows). However, when a higher InnerIter2 = $20$ was applied, the misalignment around the diaphragm (Fig.~\ref{fig:fig2}, left column, bottom row) was still observed after $100$ outer iterations. In contrast to Approach I, the performance of the misalignment estimation of Approach II was improved as InnerIter2 increased (Fig.~\ref{fig:fig2}, right column). When InnerIter2 $\geq 10$ was used, the algorithm was able to realign the input $\bm{\mu}$ map to the target one at OuterIter = $100$. 

\begin{figure}
	\centering
    \includegraphics[width=0.49\linewidth]{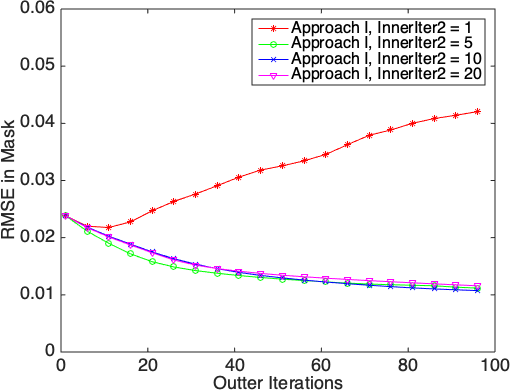}
    \includegraphics[width=0.49\linewidth]{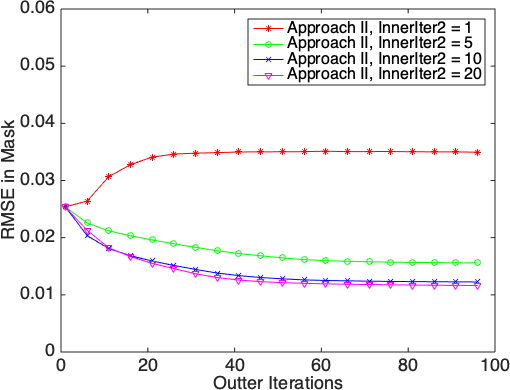}
	\caption{The RMSE in the mask plotted against the outer iteration numbers for Approach I (left) and II (right). The applied workflows were $1$ inner iteration for the misalignment estimation (InnerIter1) and $1$, $5$, $10$ or $20$ inner iterations for the image reconstruction (InnerIter2). The noiseless dataset was used.}
	\label{fig:fig3}
\end{figure}

The RMSE in the mask was consistent with the visual observation from the difference images. As shown in Fig.~\ref{fig:fig3}, for both approaches, at OuterIter = $100$, the RMSE in the mask reached the highest error when InnerIter2 = $1$ was used. For Approach I, the optimal InnerIter2 was $10$, while the RMSE in the mask decreased as InnerIter2 increased for Approach II. In terms of the convergence rate, Approach II was able to reach a relatively stable RMSE in the mask after $60$ outer iterations. In contrast, Approach I did not yet reach a stable RMSE in the lungs at $100$ outer iterations for all evaluated workflows. 

\begin{figure}
	\centering
	\includegraphics[width=0.49\linewidth]{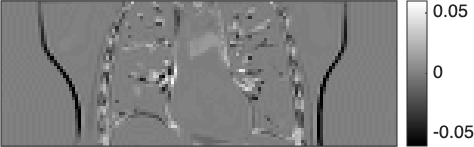}
	\includegraphics[width=0.49\linewidth]{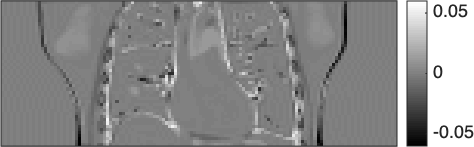}
	\includegraphics[width=0.49\linewidth]{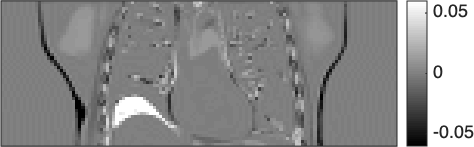}
	\includegraphics[width=0.49\linewidth]{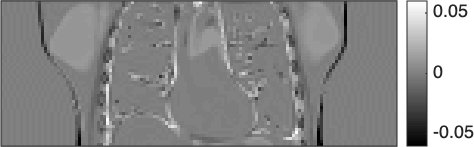}
	\includegraphics[width=0.49\linewidth]{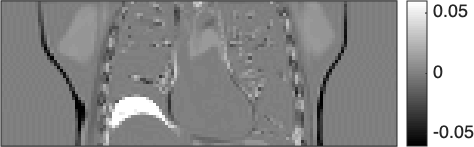}
	\includegraphics[width=0.49\linewidth]{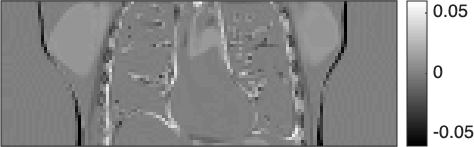}
	\includegraphics[width=0.49\linewidth]{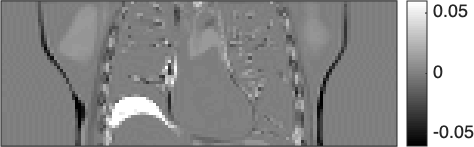}
	\includegraphics[width=0.49\linewidth]{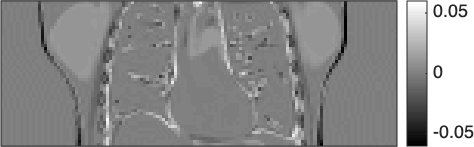}
	\caption{The central coronal view of the difference images between the target and $\bm{W}{\bm{\mu}'}$ maps for Approach I (left column) and Approach II (right column) after $100$ outer iterations. The applied workflows were $10$ inner iteration for the image reconstruction and $1$ (top row), $5$ (second row), $15$ (third row) and $30$ (bottom row) inner iterations for the misalignment estimation. The noiseless dataset was used.}
	\label{fig:fig4}
\end{figure}

Since both approaches provided visually and numerically good results when InnerIter2 = $10$ was used, we fixed InnerIter2 = $10$ and increased the iteration number used in the misalignment estimation subroutine. Fig.~\ref{fig:fig4} shows the central coronal view of the difference image between the warped and target $\bm{\mu}$ maps for both approaches with various InnerIter1 and a fixed InnerIter2 = $10$. The misalignment around the diaphragm region became apparent for Approach I at OuterIter = $100$ as InnerIter1 $> 1$ was chosen (Fig.~\ref{fig:fig4}, left column). For Approach II, however, the difference images at OuterIter = $100$ were visually identical, regardless of the number of the applied InnerIter1 (Fig.~\ref{fig:fig4}, right column). 

\begin{figure}
	\centering
    \includegraphics[width=0.49\linewidth]{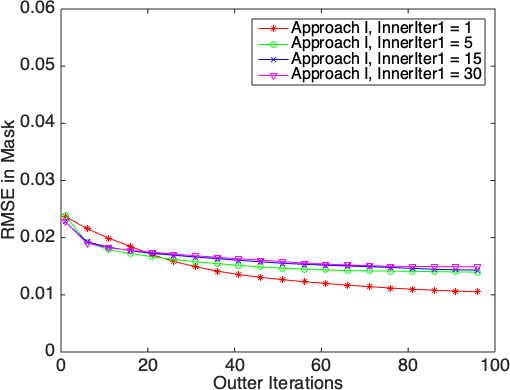}
    \includegraphics[width=0.49\linewidth]{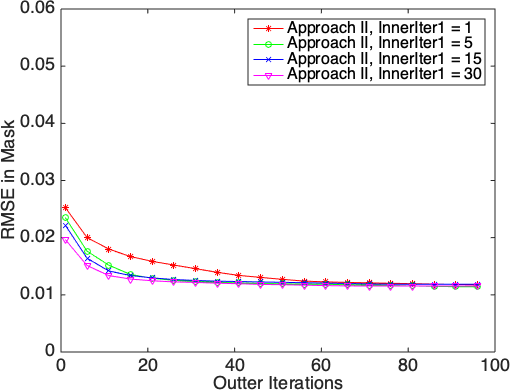}
	\caption{The RMSE in the mask plotted against the outer iteration numbers for Approach I (left) and II (right). The applied workflows were $10$ inner iteration for the image reconstruction (InnerIter2) and $1$, $5$, $15$ or $30$ inner iterations for the misalignment estimation (InnerIter1). The noiseless dataset was used.}
	\label{fig:fig5}
\end{figure}

The RMSE in the mask for Approach I and II with different InnerIter1 plotted against the outer iteration numbers are given in Fig.~\ref{fig:fig5}. Consistent with the visual comparison, for the first approach, the smallest RMSE in the mask at OuterIter = $100$ was achieved by the workflow with InnerIter1 = $1$. When Approach II was adopted, all workflows with different InnerIter1 settings were able to achieve a similar RMSE in the mask at OuterIter = $100$. The convergence rate of the misalignment estimation of Approach II was improved as InnerIter1 increased. However, the performance of Approach I in terms of the convergence rate of the RMSE in the mask seemed insensitive to the change of the inner iteration number for the misalignment estimation when the applied InnerIter1 was larger than $1$.   

\begin{figure}
	\centering
	\includegraphics[width=0.49\linewidth]{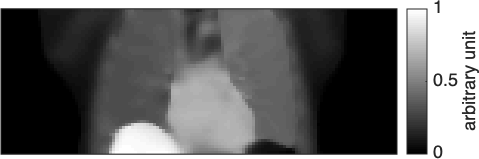}
	\includegraphics[width=0.49\linewidth]{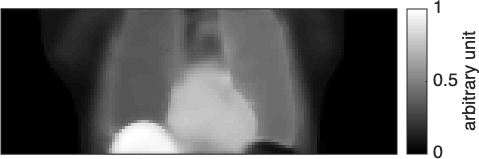}
	\includegraphics[width=0.49\linewidth]{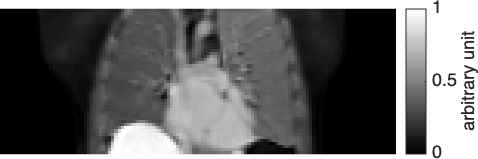}
	\includegraphics[width=0.49\linewidth]{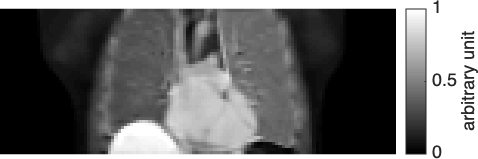}
	\includegraphics[width=0.49\linewidth]{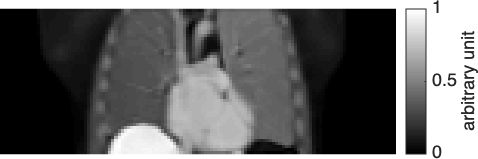}
	\includegraphics[width=0.49\linewidth]{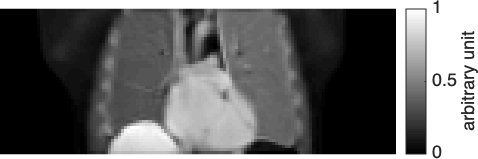}
	\caption{The central coronal view of the activity images at $100$ outer iterations (top row) and after an additional reconstruction with a smaller $\beta$ (second row) for Approach I (left column) and II (right column). The target activity images for the case where well-aligned $\mu$ map and a smaller $\beta$ are used (bottom row) are also provided.} The applied workflow was $1$ inner iteration for the misalignment estimation and $10$ inner iterations for the image reconstruction. The noiseless dataset was used.
	\label{fig:fig6}
\end{figure}

Based on the results shown in this section, $1$ iteration of misalignment estimation (InnerIter1 = $1$), followed by $10$ iterations of image reconstruction (InnerIter2 = $10$) was defined as the workflow that provides satisfactory results for these two approaches when the noiseless dataset is considered. The corresponding central coronal view of the reconstructed functional images at OuterIter = $100$ are provided in Fig.~\ref{fig:fig6} (top row). Note that the anatomical information was utilized throughout this two-part study. As shown in the figure, the selection of each penalty strength based on the performance of misalignment estimation has led to over-regularized functional images. Therefore, an additional reconstruction (using the same framework as the image reconstruction subroutine of Approach I) with a smaller $\beta = 7\times10^{-5}$ was also performed following the last outer iteration (Fig.~\ref{fig:fig6}, second row). To provide insights into the relevance between the realigned and target activity images when clinically appropriate settings for the image reconstruction are used, the estimated activity images using well-aligned $\mu$ map for calculating the attenuation correction factors and anatomical prior are given as well (Fig.~\ref{fig:fig6}, bottom row).

\subsection{Influence of availability of anatomical information}
\begin{figure}
	\centering
	\includegraphics[width=0.49\linewidth]{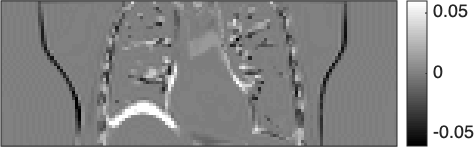}
	\includegraphics[width=0.49\linewidth]{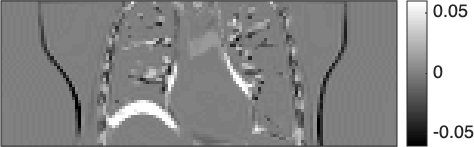}
	\includegraphics[width=0.49\linewidth]{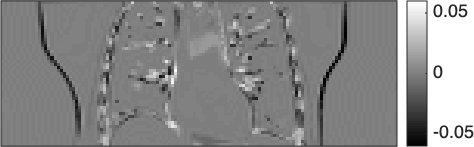}
	\includegraphics[width=0.49\linewidth]{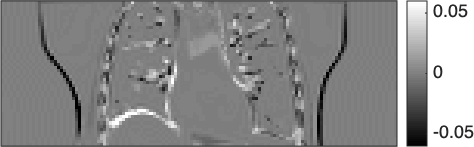}
	\includegraphics[width=0.49\linewidth]{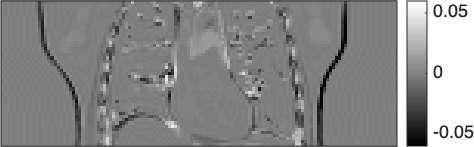}
	\includegraphics[width=0.49\linewidth]{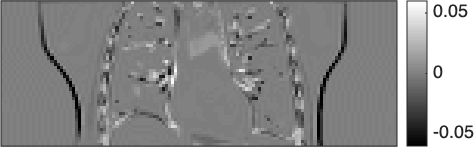}
	\caption{The central coronal view of the difference images between the target and $\bm{W}{\bm{\mu}'}$ maps for Approach I at $20$ (top row), $60$ (second row) and $100$ (bottom row) outer iterations. The results for the reconstructions without and with considering the anatomical information are shown in the left and right column, respectively. The noiseless dataset was used.}
	\label{fig:fig7}
\end{figure}

\begin{figure}
	\centering
	\includegraphics[width=0.49\linewidth]{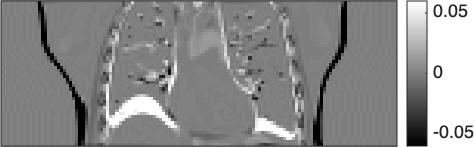}
	\includegraphics[width=0.49\linewidth]{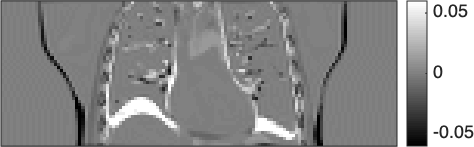}
	\includegraphics[width=0.49\linewidth]{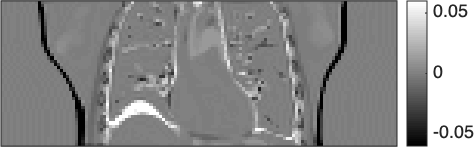}
	\includegraphics[width=0.49\linewidth]{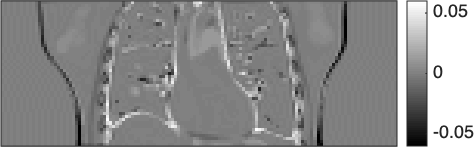}
	\includegraphics[width=0.49\linewidth]{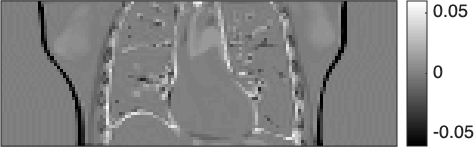}
	\includegraphics[width=0.49\linewidth]{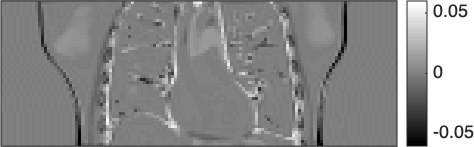}
	\caption{The central coronal view of the difference images between the target and $\bm{W}{\bm{\mu}'}$ maps for Approach II at $20$ (top row), $60$ (second row) and $100$ (bottom row) outer iterations. The results for the reconstructions without and with considering the anatomical information are shown in the left and right column, respectively. The noiseless dataset was used.}
	\label{fig:fig8}
\end{figure}

\begin{figure}
	\centering
    \includegraphics[width=0.49\linewidth]{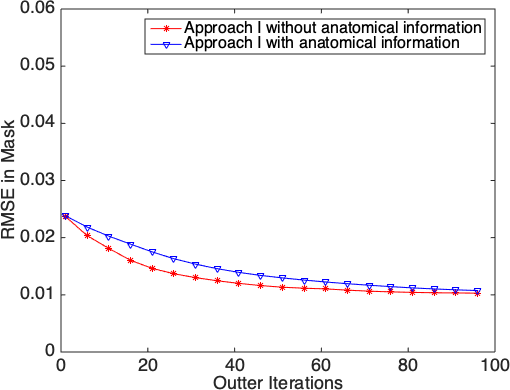}
    \includegraphics[width=0.49\linewidth]{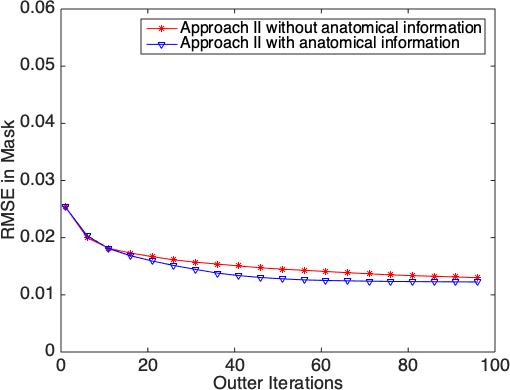}
	\caption{The RMSE in the mask plotted against the outer iteration numbers for Approach I (left) and II (right) without and with considering the anatomical information. The applied workflow was $1$ inner iteration for the misalignment estimation and $10$ inner iterations for the image reconstruction. The noiseless dataset was used.}
	\label{fig:fig9}
\end{figure}

As observed in the central coronal view of the difference images at $100$ outer iterations (Fig.~\ref{fig:fig7}~and~\ref{fig:fig8}, bottom row), both Approach I and II were able to estimate the misalignment and warp the input attenuation map accordingly, regardless of the presence of the anatomical information. However, in terms of the convergence rate, these two approaches had different responses to the use of the anatomical information. For the first approach, incorporating the anatomical information degraded the convergence rate of the misalignment estimation. The central coronal view of the difference images at OuterIter = $20$ and $60$ for the reconstructions without using anatomical information showed less apparent misalignment around the diaphragm region compared to those for the reconstructions considering the anatomical information (Fig.~\ref{fig:fig7}, top and second rows). In contrast, Approach II was able to achieve a faster convergence rate when the additional anatomical information was available (Fig.~\ref{fig:fig8}, top and second rows). These observations were further demonstrated by the RMSE in the mask plotted against the outer iteration numbers (Fig.~\ref{fig:fig9}). When the anatomical information was considered, Approach II reached a stable RMSE in the mask after around $60$ outer iterations, while Approach I required more than $80$ outer iterations to achieve that.

\subsection{Influence of data noise level}
\begin{figure}
	\centering
	\includegraphics[width=0.49\linewidth]{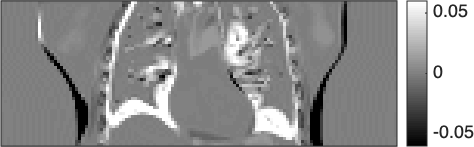}
	\includegraphics[width=0.49\linewidth]{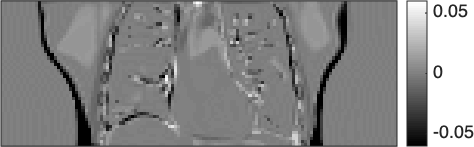}
	\includegraphics[width=0.49\linewidth]{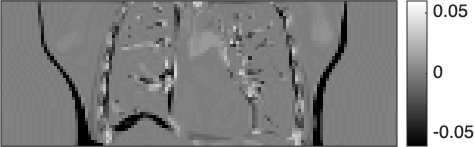}
	\includegraphics[width=0.49\linewidth]{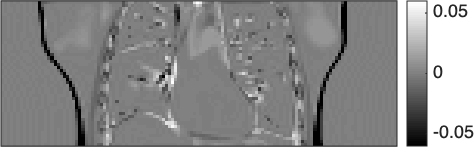}
	\includegraphics[width=0.49\linewidth]{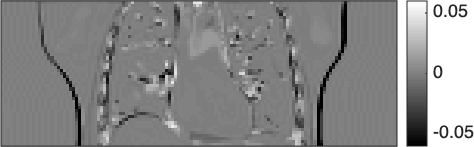}
	\includegraphics[width=0.49\linewidth]{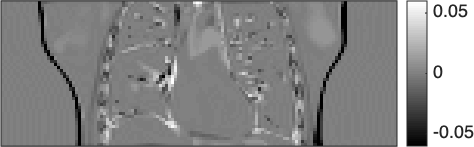}
	\caption{The central coronal view of the difference images between the target and $\bm{W}{\bm{\mu}'}$ maps for Approach I (left column) and Approach II (right column) at $100$ outer iterations. The noisy dataset with total counts of $161$~M was used. The applied workflows were $1$ inner iteration for the misalignment estimation and $10$ (top row), $20$ (middle row), $40$ (bottom row) inner iterations for the image reconstruction.}
	\label{fig:fig10}
\end{figure}

Fig.~\ref{fig:fig10} shows the central coronal view of the difference image between the warped and target $\bm{\mu}$ maps for each reconstruction condition at OuterIter = $100$ for the noisy dataset with total counts of $161$~M. As observed in the figure, the first approach still suffered from the misalignment issue at $100$ outer iterations when InnerIter2 = $10$ or $20$ were applied (Fig.~\ref{fig:fig10}, left column, top and second images). Consistent with the results for the noiseless dataset (Fig.~\ref{fig:fig2}), the performance of the misalignment estimation of the second approach was less sensitive to the applied number of InnerIter2. We obtained satisfactory results at $100$ outer iterations for both approaches using $1$ iteration of misalignment estimation, followed by $40$ iterations of image reconstruction.

\begin{figure}
	\centering
	\includegraphics[width=0.49\linewidth]{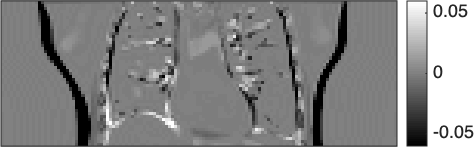}
	\includegraphics[width=0.49\linewidth]{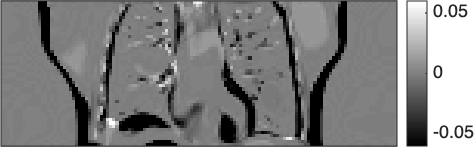}
	\includegraphics[width=0.49\linewidth]{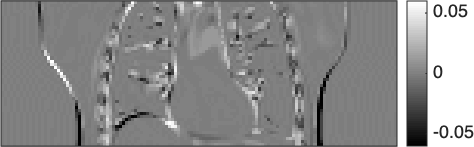}
	\includegraphics[width=0.49\linewidth]{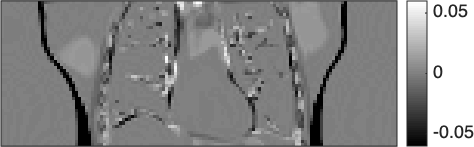}
	\includegraphics[width=0.49\linewidth]{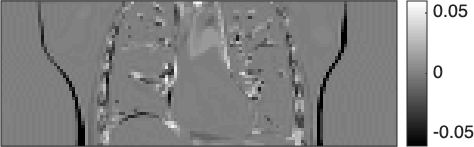}
	\includegraphics[width=0.49\linewidth]{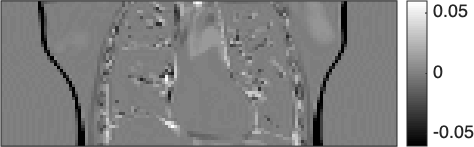}
	\caption{The central coronal view of the difference images between the target and $\bm{W}{\bm{\mu}'}$ maps for Approach I (left column) and (right column) Approach II at $100$ outer iterations. The noisy datasets with total counts of $16$~M (top row), $54$~M (second row) and $161$~M (bottom row) were used. The applied workflows were $1$ inner iteration for the misalignment estimation and $40$ inner iterations for the image reconstruction.}
	\label{fig:fig11}
\end{figure}

\begin{figure}
	\centering
    \includegraphics[width=0.49\linewidth]{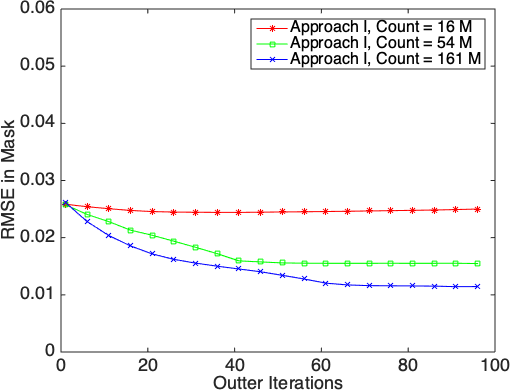}
    \includegraphics[width=0.49\linewidth]{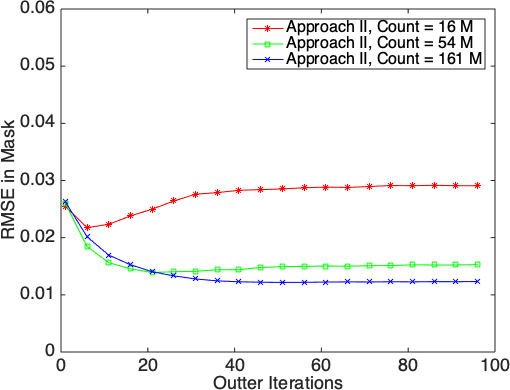}
	\caption{The RMSE in the mask on noisy data plotted against the outer iteration numbers for Approach I (left) and II (right). The applied workflows were $1$ inner iteration for the misalignment estimation and $40$ inner iterations for the image reconstruction.}
	\label{fig:fig12}
\end{figure}

\begin{figure}
	\centering
    \includegraphics[width=0.49\linewidth]{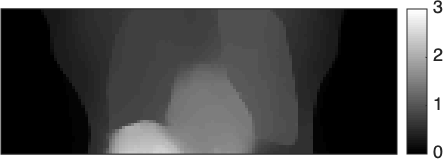}
    \includegraphics[width=0.49\linewidth]{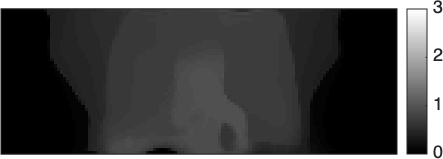}
    \includegraphics[width=0.49\linewidth]{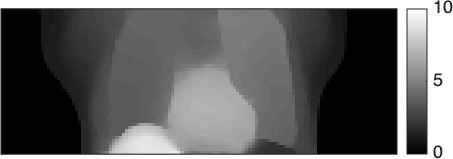}
    \includegraphics[width=0.49\linewidth]{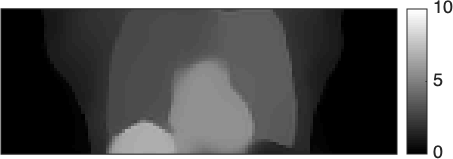}
    \includegraphics[width=0.49\linewidth]{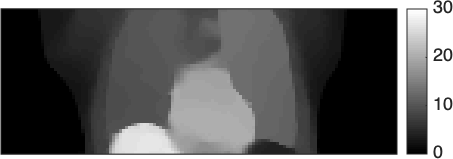}
    \includegraphics[width=0.49\linewidth]{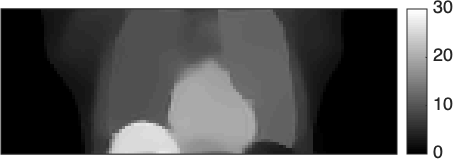}
	\caption{The central coronal view of the activity images for Approach I (left) and Approach II (right) at $100$ outer iterations. The noisy datasets with total counts of $16$~M (top row), $54$~M (second row) and $161$~M (bottom row) were used. The applied workflows were $1$ inner iteration for the misalignment estimation and $40$ inner iterations for the image reconstruction.}
	\label{fig:fig13}
\end{figure}

\begin{figure}
	\centering
    \includegraphics[width=0.49\linewidth]{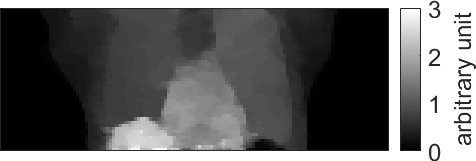}
    \includegraphics[width=0.49\linewidth]{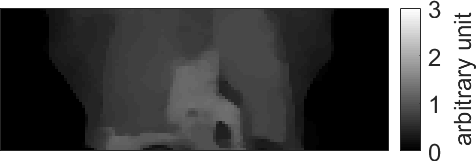}
    \includegraphics[width=0.49\linewidth]{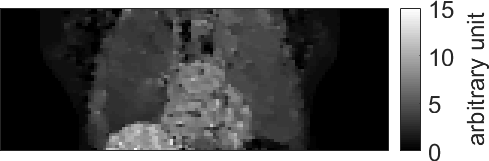}
    \includegraphics[width=0.49\linewidth]{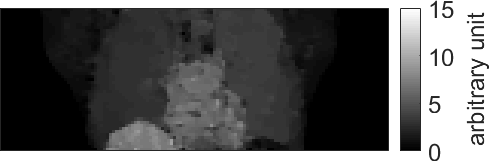}
    \includegraphics[width=0.49\linewidth]{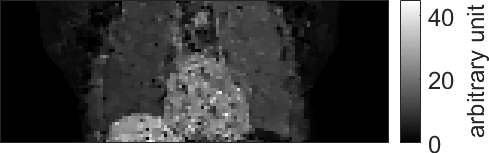}
    \includegraphics[width=0.49\linewidth]{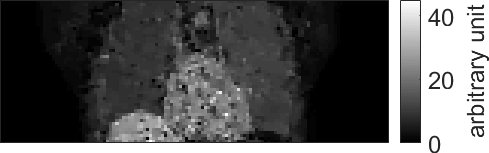}
	\caption{The central coronal view of the activity images for Approach I (left) and Approach II (right) after the additional image reconstruction. The noisy datasets with total counts of $16$~M (top row), $54$~M (second row) and $161$~M (bottom row) were used. The applied workflows were $1$ inner iteration for the misalignment estimation and $40$ inner iterations for the image reconstruction.}
	\label{fig:fig14}
\end{figure}

The central coronal view of the difference image between the warped and target $\bm{\mu}$ maps for each noisy dataset at OuterIter = $100$ is shown in Fig.~\ref{fig:fig11}. The common workflow found using the dataset with total counts of $161$~M was applied to all noisy datasets. When the data with $16$~M total count was considered, significant misalignment was still observed for both approaches at $100$ outer iterations (Fig.~\ref{fig:fig11}, top row). The second approach seemed to be more sensitive to high noise as the misalignment around the diaphragm and heart regions in the difference image was more apparent for the second approach compared to that for the first approach. The performance of the misalignment estimation was improved as the data noise level decreased for both approaches (Fig.~\ref{fig:fig11}, second and bottom row). The RMSE in the mask plotted against the outer iteration numbers support our observations from the difference images (Fig.~\ref{fig:fig12}). For data with high noise, Approach I and Approach II stuck at a high RMSE in the mask at $100$ outer iterations. When the data noise level was lower, both of them were able to achieve a lower RMSE in the mask. Moreover, Approach II converged faster than Approach I for data with low noise. The corresponding reconstructed functional images are provided in Fig.~\ref{fig:fig13}. The misalignment around the diaphragm and heart led to distortions in the activity image for the high noise dataset when Approach II was applied (Fig.~\ref{fig:fig13}, top row). Similar to the results for the noiseless dataset, the activity images were over-regularized due to the use of settings optimized for the misalignment estimation. An additional reconstruction (using the same framework as the image reconstruction subroutine of Approach I) with a smaller $\beta = 2\times10^{-2}$ was also performed. The final images are given in Fig.~\ref{fig:fig14}.

%% file: discussion.tex
\section{Discussion}
The potential misalignment between functional and anatomical images is the main concern for incorporating an anatomical prior into image reconstruction. Expanding on the algorithm proposed in \cite{Bousse2016}, two approaches that perform alternating misalignment estimation and penalized image reconstruction using anatomical priors are proposed. In this study, we focused on the performance of the misalignment estimation of each approach for the thorax with one gate of non-TOF PET data in which no motion was assumed. The $\bm{\mu}$ map was used for attenuation correction and anatomical prior calculation. Therefore, misalignment affects both factors, with subsequent influence on the optimization. Both approaches have shown the ability to estimate the misalignment and warp the anatomical image accordingly, but with a different convergence rate, depending on the applied workflow, data noise level and if the anatomical information is included as a prior. 

\subsection{Performance comparison in terms of the convergence rate of the misalignment estimation}
To study how the change of the workflow and the presence of anatomical information influence the performance of the proposed approaches, initial evaluations on noiseless datasets were conducted. As the maximum outer iteration number that controls the repetition of the alternating process was set to $100$, the workflow of both approaches was determined by the number of iterations for the misalignment estimation (InnerIter1) and image reconstruction (InnerIter2) subroutines. Based on the results shown in Section~\ref{best_work_flow}, when a sufficient number of image reconstruction subroutine was applied, Approach II showed the ability to achieve a good alignment in less outer iterations compared to Approach I. Its performance was also less sensitive to the change of workflow. The use of a larger InnerIter1 or InnerIter2 improved the convergence rate of the misalignment estimation for Approach II but slowed it down for Approach I. For the given workflow with InnerIter1 = $1$ and InnerIter2 = $10$, Approach II required a larger outer iteration to achieve a good result (Fig.~\ref{fig:fig8}) in the case where the anatomical information was absent. However, the convergence rate of misalignment estimation was improved for Approach I when the anatomical information was not available (Fig.~\ref{fig:fig7}). 
For noisy data, we found that an increased number of InnerIter2 was preferable. When the data noise level was reasonably low, both approaches showed the ability to achieve satisfactory misalignment estimation at $100$ outer iterations with the given workflow. However, for data with high noise, the proposed approaches tended to converge to a high RMSE in the mask (Fig.~\ref{fig:fig12}), indicating that they might suffer from the local minimum problem. Since it is hard to attribute the cause to the high noise level or the applied workflow, further investigation on the need of optimizing the workflow for different data noise level is necessary.

\subsection{Performance comparison in terms of the RMSE value in the mask and computation time}
Throughout the study, the RMSE in the mask was used to evaluate how the convergence of each approach depended on different factors. When using this metric to compare both methods, in most cases similar RMSE in the mask could be obtained by both approaches, with a slightly lower RMSE in the mask observed for Approach I in Fig.~\ref{fig:fig9}. However, the RMSE in the mask might not be sufficient to evaluate the performance superiority of one approach over the other, and an application-dependent metric would be advisable. In addition, it is likely that a more thorough optimization of the workflow and parameters would allow improving both methods.

In this study, a workflow that gave good results for both approaches was utilized to be able to compare their performance at a given outer iteration. Since one iteration of motion estimation is computationally more expensive than one iteration of image reconstruction, a measure that reflects the computational demand of each approach is therefore required. Although the computation time might be able to serve the purpose, the current implementation of both approaches was not optimized to have a fair comparison. They were implemented using a combination of C, C++, Fortran and MATLAB without applying acceleration techniques, such as multithreading or graphics processing unit (GPU) computing. Under the current implementation structure, Approach I and Approach II took respectively $56$ and $260$ seconds to complete one outer iteration of the suggested workflow on a desktop with an Intel i$7$-$8700$ CPU. Given that similar number of forward and back projections was involved in both approaches for the applied workflow, the difference between them in computation time could be caused by the non-optimized implementation. Future work includes further investigation and optimization of the performance of the approaches in terms of computation time.

\subsection{Influence of the strength of each penalty function}
For both approaches, a spatially-variant penalization scheme proposed in \cite{Tsai2019} was applied to the anatomical penalty function to achieve uniform local contrast. Based on the bias-variance analysis presented in \cite{Tsai2019}, the use of a stronger $\beta$ can lead to a lower image variance but higher bias, regardless of the consideration of the spatially-variant penalization. In this study, the strength of $\gamma$ and $\delta$ did not have a direct impact to the image quality as they were used to control the smoothing penalty on the motion field and barrier function on negative image values. However, since the performance of the misalignment estimation of both approaches varies with the strength of each penalty function, the image quality can be influenced indirectly by the residual misalignment caused by insufficient inner and outer iterations for a given set of hyper-parameters. We have chosen each penalty strength based on the performance of the misalignment estimation of the proposed approaches with respect to a limited set of candidate values. As can be seen in Fig.~\ref{fig:fig6} and Fig.~\ref{fig:fig13}, optimizing for accurate alignment led to over-regularized functional images. This could be overcome by running a final image reconstruction incorporating warping with settings, especially the strength of $\beta$, optimized for the functional image estimation. Example reconstructions can be found in Fig.~\ref{fig:fig6} and Fig.~\ref{fig:fig14}.

\subsection{Potential strategies to further improve the performance}
Since the deformation of the attenuation map is likely to lead to a different scatter distribution, the estimated background events should be updated accordingly during the optimization. However, the effect was assumed to be small and ignored for simplicity in the current study. To achieve accurate quantification, performing active scatter correction based on the update of the estimated activity distribution might be necessary. In practice, this could be done by re-estimating the scatter after a number of iterations of the current algorithm.

In the current study, we have adapted the strategy often used in CT or MR derived attenuation correction that down-samples the anatomical image to match the resolution of PET and surrenders the structural details carried by the high-resolution anatomical image. However, it could be beneficial to reconstruct the PET image at the same voxel size as the anatomical image instead \cite{Belzunce2017,Belzunce2019}. The benefit of using TOF data for misalignment estimation was studied in \cite{Bousse2016b} where a similar algorithm was applied to obtain reconstructed activity images with aligned attenuation correction. As Approach I and II are extensions of that algorithm, practical convergence of misalignment estimation in less outer iterations can be expected when TOF data are available.

\subsection{Future demonstrations in more realistic scenarios}
The algorithms proposed in this study have been demonstrated with simulation. A thorough validation with more realistic data is required to demonstrate usefulness in future applications in clinic. In addition, as the algorithm performance and the quantitative accuracy can be affected by other parameters that determine the edge-preserving property of the penalty (\textit{e.g.}, $\epsilon$ and $\eta$ in PLS), future work should also include parameter optimization with respect to different applications. It remains to be investigated if the introduction of the anatomical prior would introduce errors in the alignment estimation for cases where edges in functional and anatomical images are not consistent. To investigate the effectiveness of applying the proposed approaches in improving quantitative accuracy, evaluations using phantoms with inserted features or patient datasets with pseudo-lesions should also be included. In our current work, we assumed that the attenuation correction and anatomical image used for PLS were the same. However, the algorithms can be generalized to other user cases where the anatomical image is independent of the attenuation image. One particular case of this might be reconstructions of PET/CT data using an anatomical prior derived from MR images.

\subsection{Potential limitations}
The methods described here are applicable to any differentiable deformation models that can be parameterized using a linear sum of basis-functions. As in \cite{Bousse2016}, we have chosen to use the uniform cubic B-splines to model the image deformation, taking advantage of no \textit{a priori} knowledge regarding the misalignment is required. Although the model should be able to adapt to various motion effects, including respiration and organ contraction and expansion, it might fail to accurately model sliding motion such as occurs between the lower lung and ribcage. Moreover, as the current model can only address smooth motion, it is not able to handle misalignment caused by abrupt motion, for example, objects moving in and out the field of view or sudden repositioning of the arms. To improve the accuracy of the misalignment estimation and make maximum use of the anatomical information, the investigation of other valid (\textit{i.e.}, differentiable and parametrizable) deformation models specific to different applications and motion effects should be included in future work.

%% file: conclusion.tex
\section{Conclusions}
Two approaches for solving the potential misalignment between the functional and anatomical images in penalized image reconstruction using anatomical priors have been proposed in this study. The main difference between them is that the first approach deforms the anatomical image to align it with the functional one, while the second approach deforms both images to align them with the measured data. Both approaches were implemented using alternation between misalignment estimation and image reconstruction. The results demonstrated that both methods are able to estimate the misalignment and deform the anatomical image when the data noise level is reasonably low and a proper workflow for the alternating optimization is applied. Moreover, the second approach shows the ability to converge to the correct alignment faster than the first approach and is less sensitive to variations in the workflow. The use of anatomical information improves the convergence rate of misalignment estimation for the second approach, although slowed it down for the first approach. These encouraging results indicate that it is possible to align functional and anatomical information, overcoming a serious limitation in practical use of anatomical priors.